\def\year{2020}\relax
\documentclass[letterpaper]{article} 
\usepackage{aaai20}  
\usepackage{times}  
\usepackage[normalem]{ulem}
\usepackage{helvet} 
\usepackage{courier}  
\usepackage[hyphens]{url}  
\usepackage{graphicx} 
\usepackage{url}
\usepackage{graphicx}  
\usepackage[caption=false]{subfig}
\usepackage{color}
\usepackage[normalem]{ulem}
\usepackage{comment}
\usepackage{tcolorbox}
\usepackage{soul}

\usepackage[backend=biber,style=numeric,sorting=none]{biblatex}
\title{Longitudinal change in language behaviour during protests: A case study of Euromaidan in Ukraine}
\author{Ivan Slobozhan$^{1}$, Tymofii Brik$^{2}$, Rajesh Sharma$^{1}$ \\
\mbox{}\\
$^1$Insitute of Computer Science, University of Tartu, Estonia\\
$^2$ Policy Research Department, Kyiv School of Economics, Ukraine\\
\{ivan.slobozhan, rajesh.sharma\}@ut.ee, tbrik@kse.org.ua}

\begin{document}
\maketitle

\begin{abstract}
In the last decade, online social media has become the primary platform for protesters to organize and express their agenda in various parts of the world. Nevertheless, scholars still debate whether online tools induce offline protests or facilitate them \cite{shirky2011gladwell, enikolopov2020social}. Unfortunately, studies of protests often lack panel data and cannot address how particular users change their behaviour over time in line with the protest agenda. To this end, we analyze a new dataset of the Facebook page \emph{EuroMaydan} that was explicitly created to facilitate the protest in Ukraine from November 2013 to February 2014. Moreover, our analysis follows this page even after the end of the protest till June 2014. This page had more than 300,000 subscribers at the time when the data was collected. On average, this page generated hundreds of posts and thousands of comments per day, reaching up to 37,307 comments in one day. In total, the dataset includes 26,631 posts and  1,470,593 comments that were generated by 124,790 users during and after the protests. We use this panel data to test how particular users switch between the two most popular languages of this page: Ukrainian and Russian. Previous studies have puzzled with language behaviour during the Ukrainian protest \cite{etling2014russia, metzger2016tweeting}. Although researchers expected to see more Ukrainian language due to national mobilization, studies discovered that Ukrainian protesters frequently used Russian, especially after the end of the protest. A hypothesis was suggested that protesters use language strategically depending on circumstances (e.g., to maximize outreach). However, previous studies relied on aggregated data and did not explore the within-subject variation of language behaviour. Our study adds to this scholarship by adding a longitudinal analysis of language behaviour. Considering the broader contribution of this paper, we add to the literature on protests by validating previous findings derived from surveys that activists change their behaviour to reflect prior preferences and situational goals rather than modify their preferences. 
\end{abstract}

\section{Introduction}
From the most recent events in Belarus, \cite{warInst} to protests in Russia, Ukraine, Egypt, and China \cite{acemoglu2018power, dickinson2014prosymo,enikolopov2020social, onuch2015euromaidan, metzger2016tweeting, wolfsfeld2013social}; from the Occupy Wall Street movement in the US to the Indignados in Spain, to the 2015 Charlie Hebdo protest in Paris \cite{anduiza2014mobilization, theocharis2015using, larson2019social}, contemporary political protests rely on online social media (OSM), including Facebook, Instagram, Twitter, Vkontakte, and Telegram. Despite the popularity of this phenomenon in public culture and social science \cite{shirky2011gladwell}, there are still many gaps in knowledge about the dynamics of online protests. 

Most of the existing studies agree that online activities coincide with protest events. Some argue that online activities induce protests. Others suggest that online platforms exacerbate prior political preferences. However, most of the research is focused on aggregated data of protest and online events \cite{enikolopov2020social, brantly2019cyberspace, etling2014russia}. At the same time, much less attention is given to the individual-level analysis. 

Consider an example of a person who reads online about the regime's cruelties and then decides to adopt the symbols of protesters while rejecting the symbols of the elites in power. In this example, a person switches their behaviour to be in line with the cause. People can change behaviour in many ways depending on the protest movement and the cause. Some protesters adopt specific language and narratives, others change their clothes, some people change their routine and stay at the protest camps overnight instead of going home, and some protesters change their behaviour from peaceful strikes to violence. In our case, we study the online behaviour on Facebook during the Euromaidan revolution in Ukraine. Therefore, we can analyze the individual frequency of posts and comments and their content. 

Researchers have shown that the Euromaidan revolution was affected by national sentiments and national narratives, which itself are deeply connected with the national identities of Ukrainians, their personal language preferences, and support of language policies \cite{arel2018ukraine, kulyk2011language, kulyk2019identity, metzger2016tweeting, onuch2018studying, zhuravlev2020exclusiveness}. A study of the Ukrainian Twitter, Facebook, YouTube, forums, and news sites showed that those users who posted in the Ukrainian language were far more likely to express positive views about Euromaidan compared to those who posted in the Russian language \cite{brantly2019cyberspace}. Therefore, the language choices of users (Ukrainian language versus the Russian language) signals their political preferences. Curiously, previous research of the Ukrainian Twitter has shown that the proportion of content in the Russian language increased during the Euromaidan and further political shocks of the Crimean annexation \cite{metzger2016tweeting}. It could be that the Ukrainian protesters used the Russian language strategically (perhaps to target external actors or to argue with pro-Russian activists who speak Russian). Nevertheless, previous studies addressed aggregate-level data instead of individual-level data. Therefore, longitudinal change or within-subject variation (i.e., personal preference to change language behavior over time) is still understudied. 

Our paper aims to address this gap and analyze within-subject variation in online language behavior of Facebook users during the Euromaidan revolution and several months after the end of the revolution. We analyze the data from the largest Facebook group dedicated to Euromaidan with 26,631 posts and 1,470,593 comments left by 124,790 users from 22 November 2013 until 31 May 2014. Following is the step-by-step process and results, of our analysis:
\begin{enumerate}
    \item Our statistical analysis of the Facebook data at the individual level confirms previous knowledge produced by other methods with other data sources. We also find that the usage of the Russian language increased after the end of Euromaidan. We register that more than half of the active users started to use the Russian language more often after the Euromaidan. At the same time, only 4\% of active users started to use Ukrainian more often (see Section 4). 
    
    \item We then make one step further and investigate possible explanations for this finding (see Section 5).
    
    \begin{enumerate}
        \item  First, we show that the preferences of the ``veterans'' (Users which existed or have actively participated on the group page before the end of revolution) to use Russian more often with time can be associated with the influx of new users (we call this global influence in the paper) (Section 5.1)
        
        \item We do not find any evidence that the post's language correlates with the comment's language. Therefore, we rule out the possibility that the Facebook group moderators nudged people using Russian more often (Section 5.2).
        
        \item Finally, our within-subject analysis shows that the Russian language is more sticky than the Ukrainian language. Once a person decides to post in Russian, they keep doing it from one comment to another. In contrast, people who speak the Ukrainian language do it in short sequences. They often switch between languages from one comment to another (Section 5.3). This finding confirms the previous hypothesis that the Ukrainian language was used not for ethnic or national mobilization but instead strategically depending on circumstances \cite{metzger2016tweeting}
    \end{enumerate}
\end{enumerate}

Our analysis contributes to several streams of academic scholarship. Considering methodology, researchers of protests use a wide range of tools to register an actual change in individual behaviour. Some scholars employ ethnographic methods or qualitative interviews to address individual changes in language, gestures, clothes, and other symbolic clues \cite{shevtsova2017euromaidan, zhuravlev2020exclusiveness, nikolayenko2018women, zelinska2017ukrainian}. Other scholars use large-scale representative surveys to confirm that public attitudes can change over time in line with the protest agenda \cite{pop2021protest}. A new wave of computational social scientists used aggregated data of Twitter posts to register changes in the language of protesters \cite{metzger2016tweeting}. Nevertheless, there was no computational study of within-subject variation in language change during the protest to the best of our knowledge. We also confirm that the Ukrainian language was used strategically during the Euromaidan protest, which is in line with recent sociological studies which show the fluidity of linguistic behaviour and national identities of Ukrainians and simultaneously challenge the idea of ethnic mobilization  \cite{arel2018ukraine,zhuravlev2020exclusiveness, surzhko2017framing}.

\section{Related Work}
This paper investigates the online behavior during the Euromaidan revolution in Ukraine, including several months after the end of the protest. The Euromaidan is well described in the literature \cite{brantly2019cyberspace,onuch2018studying, zelinska2017ukrainian, pop2021protest, metzger2016tweeting}. Sufficient to say, it was a large grassroots political movement with hundreds of thousands of Ukrainians protesting in Kyiv and other cities across the country. The protest was caused by the President of Ukraine, Viktor Yanukovych, who did not sign an association agreement with the European Union. Instead, he announced that economic ties between Ukraine and Russia would be a priority. A small group of students and young people organized a protest in the center of Kyiv. The police brutally attacked this protest, which mobilized a wide range of social groups to join the protest. New protesters criticized police brutality, the legitimacy of the regime, and the pro-Russian agenda of Viktor Yanukovych. From the very beginning, this protest was significantly affected by social media. With time, in a series of dramatic events, the protest escalated to violent clashes. Hundreds of protesters died, snipers shot many. Viktor Yanukovich fled the country to Russia. These events escalated relations between Ukraine and Russia, resulting in the annexation of Crimea and the beginning of war in Donbas.

Researches have shown the Euromaidan led to the decline in support for a close relationship with Russia, and to the increase in the proportion of people thinking of Ukraine as their homeland \cite{pop2018identity}. Other researchers also show that the number of Ukrainians who became proud of their citizenship increased after Euromaidan \cite{golovakha2020conclusions}, and that there was a shift towards national identity consolidation across different social groups and regions \cite{kulyk2019identity}. At the same time, the same research has shown that Ukrainians are more likely to shift attitudes to reflect their identities rather than modify their identities \cite{pop2018identity, peisakhin2018electoral}. This observation is complemented with the fact that average ethnic identities and language practices of Ukrainians changed little after the Euromaidan \cite{pop2018identity}. This set of findings is essential for our study for two reasons. First, Euromaidan was fundamentally tied with national identities and attitudes. At the same time, these studies show that the identities of Ukrainians were not likely to be induced by online mobilization. In contrast, it was more likely that Ukrainians were using online platforms to express their prior political beliefs. 

Researchers of the Ukrainian Twitter arrived at similar conclusions. Although some scholars expected to find an increased "ethnic mobilization" and increased shares of the Ukrainian language on Twitter, they discovered the opposite \cite{metzger2016tweeting}. They saw an increase in the proportion of the Russian language after the end of the Euromaidan. Similar findings were suggested by Etling, who showed more support for the Euromaidan protests in Russian-language sources than he initially expected \cite{etling2014russia}. A possible explanation was suggested that Ukrainians used the Russian language strategically (perhaps to target relevant audiences)\cite{metzger2016tweeting}. Other researchers suggested that the narrative of Twitter posts changed with time. It changed from describing the Euromaidan as a peaceful movement to existential danger to the Russophone population \cite{lyebyedyev2018euromaidan}. This change of the narrative was likely to be one of the driving forces of the growing proportions of the Russian language. Studies of other platforms beyond Twitter were scarce. A study by Surzhko-Harned and Zahuranec utilized only 1107  posts on Facebook and showed that the protest conceptualized their movement in terms of domestic issues and an anti-regime revolution rather than a geopolitical crossroad between the EU and Russia \cite{surzhko2017framing}. Dickinson theorized that Facebook was used for logistics of the protests \cite{dickinson2014prosymo}, and Onuch used surveys to confirm that online social media, including Facebook, were used to recruit participants \cite{onuch2015facebook}. 

Overall, computational studies indicate a general surprise among the international researchers that the proportion of the Russian language online did not diminish with time. In fact, the opposite was observed. At the same time, scholars of the history and politics of Ukraine were not so surprised because they knew for a long time that Ukrainians from different regions use both languages circumstantially (at work or at home) \cite{arel2018ukraine}. This scholarship indicates that Ukrainian identities do not necessarily correlate with language behavior, meaning that Ukrainian patriots can express their ideas in both Russian and Ukrainian.

The debate about how Ukrainian users used language during the protest online is still not conclusive. For example, Metzger et al. \cite{metzger2016tweeting} state explicitly in their paper that: \emph{“…it is also important to move beyond the aggregate level data to consider within-subject variation. This will allow us to look at how specific users change their behavior over time, and to consider alternative explanations for aggregate trend…”} (p. 51). To this date, all empirical knowledge about the Euromaidan and online language behavior is based on aggregated data. Therefore, we still do not know the extent to which users modified their preferences and changed language behavior with time.

In general, studies of online communications suggest that online platforms can influence changes in individual identities \cite{pempek2009college, klandermans2014identity}, cultural taste \cite{lewis2008tastes}, and voting attendance \cite{bakshy2012role, bond201261}. Moreover, political science theory suggests that online platforms provide unique (and otherwise absent) information about the quality of governments to people, thus increasing the likelihood chances of a person to change their behavior and join the protest with time \cite{edmond2013information, little2016communication}. In addition to this, online platforms are effective to coordinate people logistically \cite{enikolopov2020social, bennett2012logic} and engage them emotionally \cite{jost2018social}. Drawing from this scholarship, one can assume that the individual behavior during the Euromaidan should have been affected by social media. However, the exact influence is yet to be tested. Some scholars would suggest that online users should have increased their personal usage of Ukrainian because the Ukrainian language significantly correlated with a positive evaluation of Euromaidan \cite{brantly2019cyberspace}. Other scholars would suggest that individual language preferences should be volatile because they are affected by circumstances and strategic situations \cite{metzger2016tweeting}. In what follows, we will test these competing ideas using the longitudinal data of Facebook activists.

\section{Dataset}
\label{chapter:dataset}

We collect dataset from the Facebook group called \textit{EuroMaydan}\footnote{\url{https://www.facebook.com/EuroMaydan}} using the Netvizz application \cite{rieder2013studying}. 
We select Facebook platform because it is one of the most popular social networking sites in Ukraine with over 3 million active users by 13 October 2013 \footnote{\url{https://en.wikipedia.org/wiki/Internet_in_Ukraine#Facebook}}. Facebook groups provide a place to connect people with common interests. In groups, the users can share their thoughts and discuss the posts of other participants. The groups are managed by admins or moderators that have superior rights compare to a general group member. For example, admins can restrict the users' rights to make posts in the groups, so the members can only make comments to the posts. This is also the case of \textit{EuroMaydan} group, where the users have such restrictions, while only the admins or group moderators are allowed to make the posts and, as a result, shape the group agenda. There is also a difference in the set of rights between the admin and the moderator, however this is not crucial for our analysis. 
Next, we briefly discuss the difference between two main communication entities in Facebook groups: posts and comments. The main difference between posts and comments is that a post is a message to the group that starts a new thread (or discussion) in a group timeline, while a comment usually is a message or response to the post. We assume that the reader is probably familiar with Facebook so that we won't dig into details of this platform. Instead, in what follows we switch to the more detailed analysis of our dataset.

\textit{EuroMaydan} group was created on 21 November 2013 and was the largest page dedicated to the protest. The dataset we scraped included 26,631 posts and 1,470,593 comments left by 124,790 users from 22 November 2013 until 31 May 2014. Table \ref{table:column_description} shows a more detailed metadata for it. The dataset is fully anonymised and does not contain any information about a users' age, gender, place of living, list of friends, etc. However, the metadata contains users' unique (anonymised) identifiers, which helped differentiate the users, but not the users' real identities. Thus, our analysis and the findings being reported do not violate the privacy of the users. 

\begin{table}[h]
\centering
\resizebox{\columnwidth}{!}{%
\begin{tabular}{|l|l|}
\hline
\textbf{Column}    & \textbf{Description}                     \\ \hline
post\_id           & unique identifier of the post            \\ \hline
post\_by           & unique identifier of the user who published the post    \\ \hline
post\_text         & text of the post                         \\ \hline
post\_published    & date when the post was published         \\ \hline
comment\_id        & unique identifier of the comment         \\ \hline
comment\_by        & unique identifier of the user who published the comment \\ \hline
comment\_text      & text of the comment                      \\ \hline
comment\_published & date when the comment was published      \\ \hline
\end{tabular}%
}
\caption{Metadata}
\label{table:column_description}
\end{table}

\section{Did online activists change their language usage?}
\label{sec:Language switching patterns}

As we discussed in the previous chapters, some researchers suggested that online users were likely to change their language patterns because they were mobilized by crucial political events. Others suggested that users were more likely to change because after the end of the revolution they were exposed to various strategic situations (e.g., they had to argue with Russian speakers in the Russian language or to target international audiences). In this section, we analyze the users' usage of Russian and Ukrainian languages in comments. Following previous scholarship, we analyze trends before and after the end of the Euromaidan and the annexation of Crimea by the Russian Federation \cite{metzger2016tweeting}\footnote{\url{https://en.wikipedia.org/wiki/2014_Ukrainian_revolution}}

\subsection{Language classification}
Our analysis starts with an explanation of how we classify the language of comments because it is one of the pillars of our research question. 
We first preprocessed raw texts of comments using a standard text preprocessing pipeline: lowercase the text, remove punctuation, URLs, and numbers. Then to identify the language of the comment, we use a text language classification model from FastText\footnote{\url{https://fasttext.cc/}}. To verify the language prediction performance by the FastText model, we calculate the accuracy of the model on a sample of 2000 random comments from our dataset. We manually label each of the comments from the sample and then compare the label with the models' prediction.
It turns out that the accuracy is 96\% for the language classification task, which we consider as a good score for the reliable results and thus we continue our analysis. Based on the language prediction, out of all the comments, 59\% of them are in Russian, 31\% in Ukrainian, and 10\% are comments in other languages. We then exclude comments in languages other than Russian and Ukrainian due to their marginal presences, and in addition, they are beyond the scope of our research question. 
Then, we aggregate the sorted comment languages by the date of comment on the level of user. We received a vector for each user with values either 'Russian' or 'Ukrainian'. Finally, we map these values to binary: 'Russian' to 0 and 'Ukrainian' to 1, assuming that the comment in the Ukrainian language was a success and comment in the Russian was a failure in our sample. In addition, we used alternative mapping depending on our research question, where the Ukrainian was a failure, and the Russian was a success (a more detailed explanation will be in the following section).

\subsection{The identification of the change of the language usage behaviour}
\label{subsection: language change identification}

In order to evaluate how many online activists did change their language patterns, we counted how often each of them posted in Russian and Ukrainian, respectively. Then, for each user, we performed a statistical test to compare individual proportions of each of the language before and after the end of the Euromaidan.

\begin{figure}[!h]
\centering
\includegraphics[width=\columnwidth]{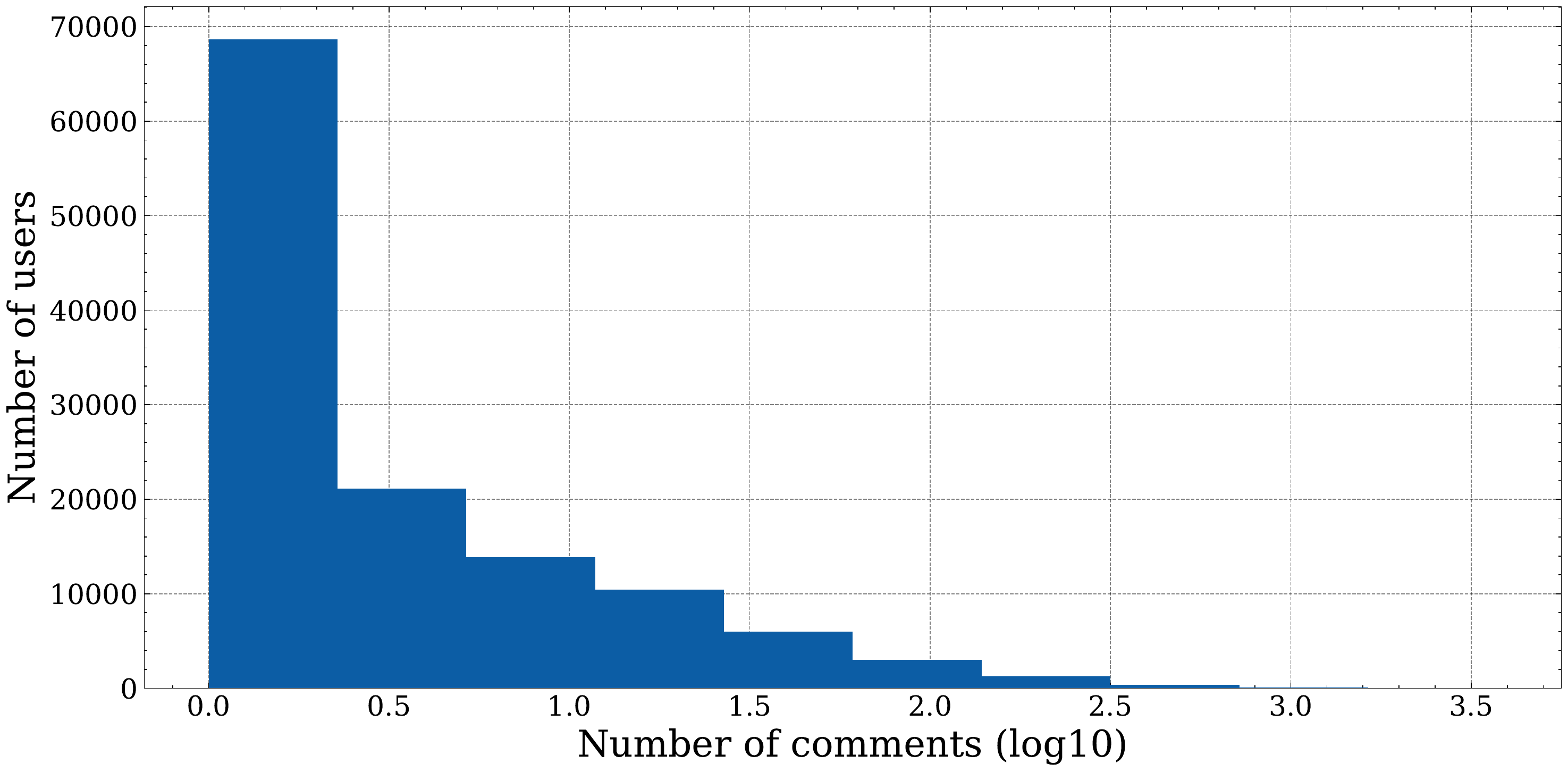}
\caption{Histogram of the number of comments per user}
\label{figure:number_of_comments_histogram}
\end{figure}

More, precisely the null hypothesis is \textit{P1} - \textit{P2} $\geq$ 0, and the alternative hypothesis is \textit{P1} - \textit{P2} $<$ 0, where \textit{P1} is the proportion of the comments in the Ukrainian language before the end of the revolution and \textit{P2} is also a proportion of comments in Ukrainian language but after the revolution. In the same manner, we also compare the proportions of Russian comments before and after the end of Euromaidan. We select one-sided hypotheses over two-sided because we are precisely interested in whether the users started to use particular language more frequently, rather than simply checking whether there has been any change or not. This statistical approach requires a sufficient number of comments per user. Otherwise, the analysis can be harmed by spurious effects due to a lack of data. As Figure \ref{figure:number_of_comments_histogram} shows that most of the users in our dataset are not active and commented only a few times. More precisely, the mean number of comments per user is 12, while the second (median) and third quartile are 2 and 7, respectively. Therefore, to select only relevant users that can provide us reliable results, we fix a \textit{threshold} by the number of their comments. We define the users who have commented for at least \textit{threshold} number of times as \textit{active users}. To find this \textit{threshold} by the number of comments, we follow the suggestion from Agresti and Franklin \cite{agresti2007art}. The authors propose a rule of thumb for its minimum sample size: at least ten successes and ten failures in each sample. In our case, it means that the user should have at least ten comments in the Russian language and ten comments in the Ukrainian language in one sample (before the split date) and another (after the split date).

\subsubsection{Robustness check: }

We run robustness checks to make sure that our findings are stable across different thresholds. To this end, we marginally increase our requirement to the minimal sample size in several steps. Firstly, we select those users with ten comments in the Russian and ten comments in the Ukrainian language (as suggested above) before and after the split date (22 February 2014, the end of Euromaidan). This resulted in 557 users. In addition, we also try a higher threshold: 15 comments in the Russian and 15 comments the Ukrainian, which resulted in 337 users. 
Furthermore, we vary our split date, which initially indicates the end of the protest. We repeat our analysis with the split date of one month after (marked as + 1 month) and two months after (marked as + 2 months) the Euromaidan. We do not select the earlier date because we don't have a lot of data one month before the Euromaidan, as our dataset includes comments from roughly the end of November 2013.
As a part of the sanity check that our approach is correct, we expect to observe:
\begin{enumerate}
    \item The higher \textit{threshold} should not significantly alter the proportion of rejections of the null hypothesis. 
    \item The number of rejections of the null hypotheses should be lower if we choose either one month or two months after the date of the Euromaidan revolution as a split date. This is because the end of Euromaidan, as we assumed, should be one of the most influential events towards Ukrainian self-identification during the period we analyzed. Thus, this should be the date when most people start to change their behaviour in one way or another. Otherwise, we can expect no significant difference between the split dates if the Euromaidan hasn't altered the users' language preferences. 
\end{enumerate}

\subsubsection{Findings: }
We calculate the number of rejections of the null hypothesis for each of the \textit{active user}, depending on the \textit{threshold} and the split date. In all the cases, we use a significance level of 0.05. The results for the Russian and the Ukrainian languages can be found in Table \ref{table:Russian results} and \ref{table:Ukrainian results}, respectively.

\begin{table}[h]
\centering
\resizebox{\columnwidth}{!}{%
\begin{tabular}{|l|l|l|l|}
\hline
\textbf{}              & \textbf{Euromaidan} & \textbf{+ 1 month} & \textbf{+ 2 months} \\ \hline
\textbf{Threshold: 10} & 46\%                 & 27\%                 & 26\%                 \\ \hline
\textbf{Threshold: 15} & 52\%                 & 33 \%                 & 28 \%                 \\ \hline
\end{tabular}%
}
\caption{Percentages of users who started using Russian language more often after the split date}
\label{table:Russian results}
\end{table}

\begin{table}[h]
\centering
\resizebox{\columnwidth}{!}{%
\begin{tabular}{|l|l|l|l|}
\hline
\textbf{}              & \textbf{Euromaidan} & \textbf{+ 1 month} & \textbf{+ 2 months} \\ \hline
\textbf{Threshold: 10} & 4\%                  & 6\%                  & 7\%                  \\ \hline
\textbf{Threshold: 15} & 2\%                  & 6\%                  & 4\%                  \\ \hline
\end{tabular}%
}
\caption{Percentages of users who started using Ukrainian language more often after the split date}
\label{table:Ukrainian results}
\end{table}


We can observe that there is statistical evidence in the change of the language behaviour for the \textit{active users} and it passed our robustness check as expected. In addition, it differs for each language. Approximately half of the active users started to use Russian more after the end of Euromaidan, while only 4\% of active users started to prefer Ukrainian more.
This is very intriguing and begs the following questions: 
\begin{enumerate}
    \item Firstly, why this change occurred (for the Russian language) and are there any factors that can be related to this change? (Discussed in Sections 5.1 and 5.2)
    
    \item Another question is why did active users start using the Russian language more frequently, and so few active users started using Ukrainian more often? (Discussed in Section 5.3)
\end{enumerate}
 
 Although we cannot provide a conclusive answer to these questions, we will use our data to propose possible explanations in the following section.


\section{Possible explanations of the individual change in language behaviour}

\label{sec:Language topic patterns}
In this section, we propose some possible explanations to what we have observed in Section \ref{sec:Language switching patterns}. 
We cannot test these hypotheses directly due to the limitations of our data, and we cannot analyze the social and psychological traits of the users since we do not have access to this information (more about limitations in Section \ref{chapter:limitations}). Nevertheless, we propose several explanations of this behaviour based on the avaliable data of posts and comments in the dataset.

\subsection{Global influence: If the group shape the language preference of the active users?}

First of all, we analyze whether the composition of the group has any effect on language preference. We focus on the Russian language in this part of the analysis because, for this language, we have found strong evidence that there was a change, as approximately half of the \textit{active users} started to use the Russian language more often. One can think of two alternative scenarios. First, the Facebook page is stable. There are no new users. Therefore, any prevalence of the Russian language can be explained by the behaviour of "veteran users" who prefer to stick to this language. Second, alternatively, it could be the case that the Facebook page is dynamic, and many new users join with time. If these new users prefer to speak Russian, it is likely that they will drive the overall language behaviour. In what follows, we try to analyze this composition effect of old and new users, which we refer as \textbf{Global influence}" and remove the next line. This mechanism works in a twofold manner. First, when users are exposed to a lot of comments in the Russian language in the group, they start using it out of peer pressure. Second, some users might use the Russian language strategically in order to engage with meaningful conversations and spread their ideas. We cannot disentangle these mechanisms due to the limitation of our data, but at least we can observe the influence of the overall composition effect. 

\begin{figure}[t]
\centering
\includegraphics[width=\columnwidth]{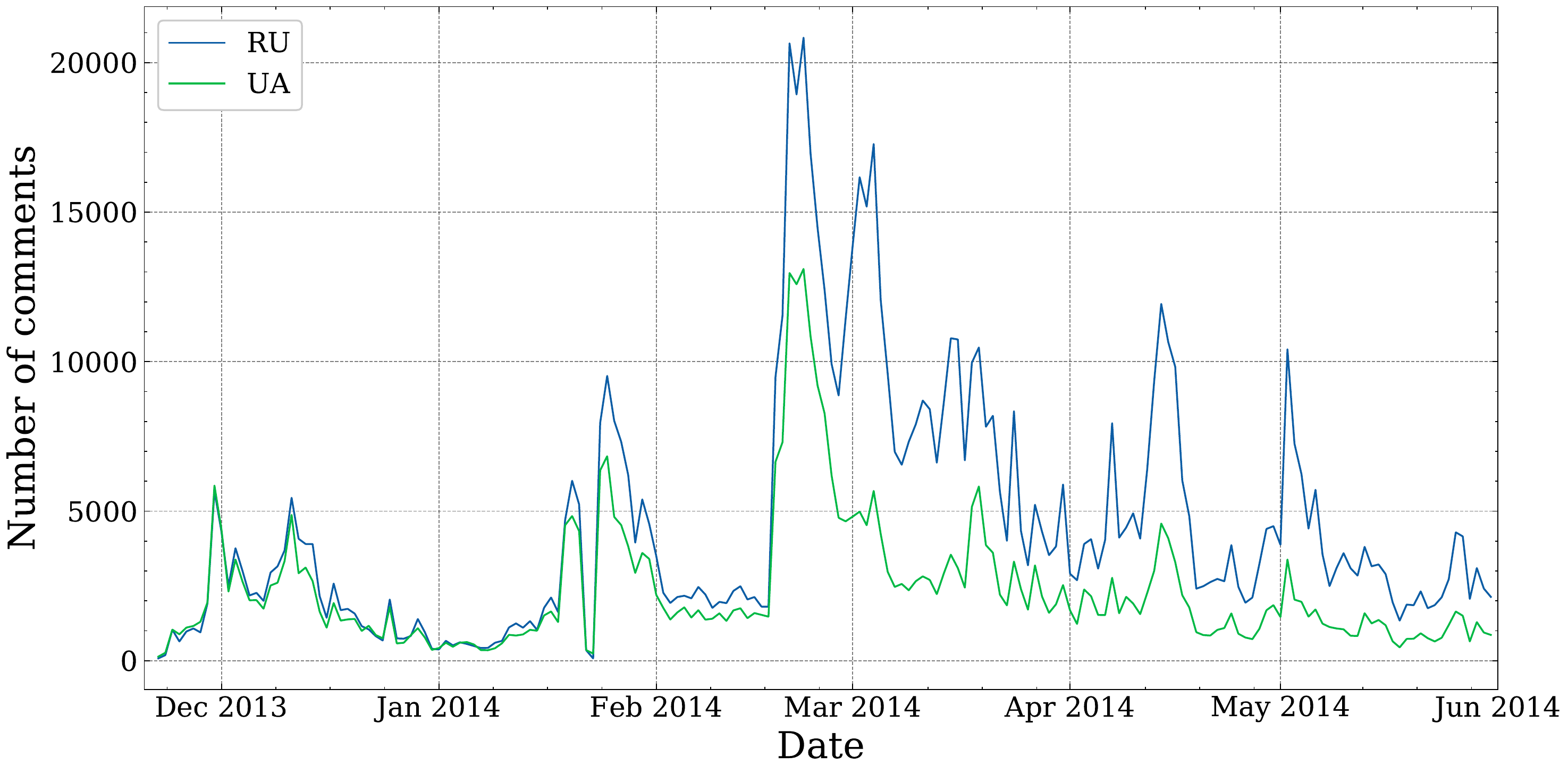}
\caption{Comments in Russian and Ukrainian language}
\label{figure:russian_and_ukrainian}
\end{figure}

\begin{figure}[t]
\centering
\includegraphics[width=\columnwidth]{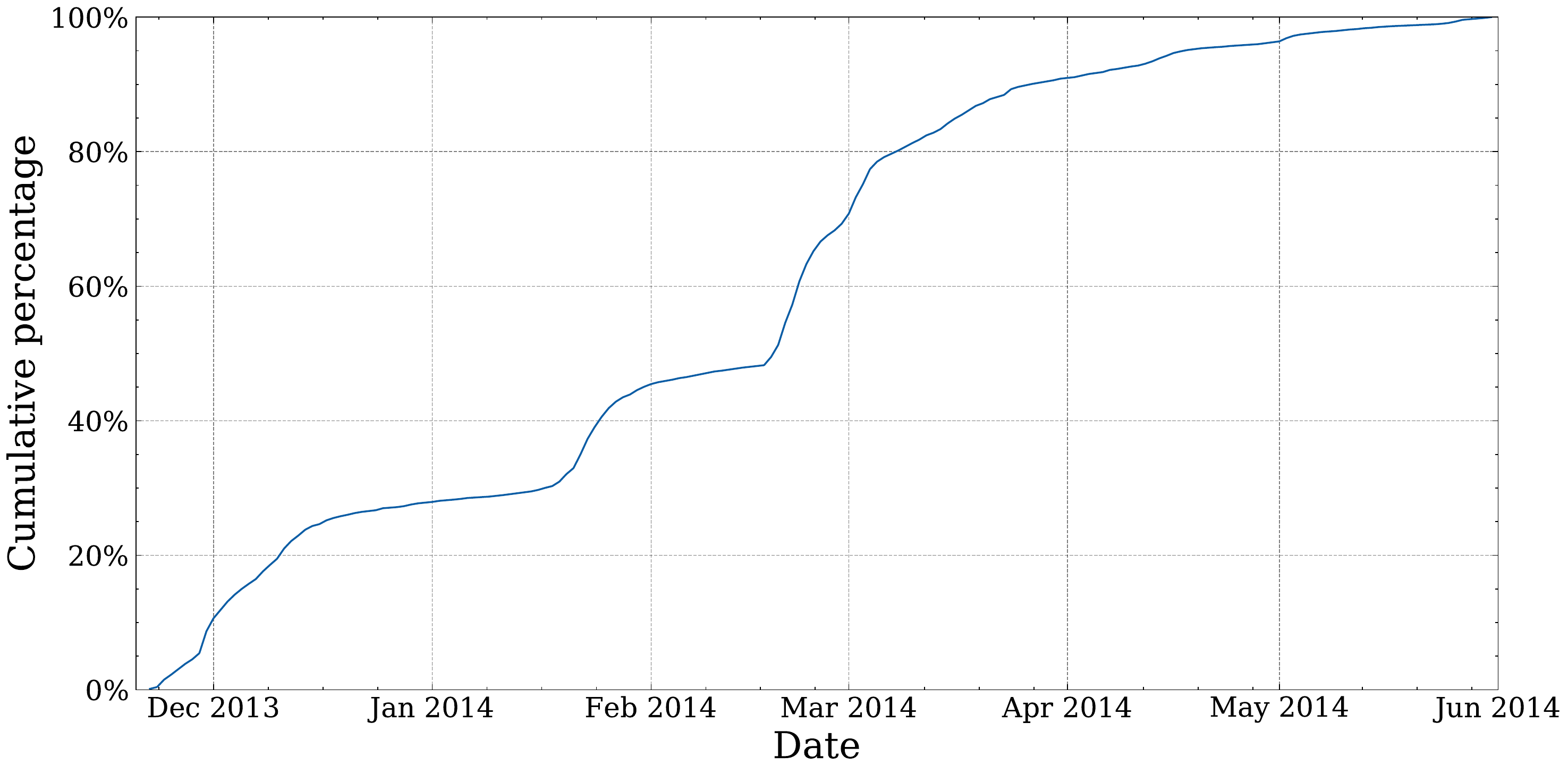}
\caption{The influx of new users}
\label{figure:new_users}
\end{figure}

Figure \ref{figure:russian_and_ukrainian} shows the patterns of using Russian and Ukrainian languages went hand in hand before February 2014 (the month in which Euromaidan ended). Then, the number of Russian comments increased, especially after March 2014. Figure \ref{figure:new_users} shows the influx of new users (we define new users as users who made their first comment on a given date) as one possible explanation of this behaviour. The influx of new users clearly coincided with the end of the Euromaidan. Therefore, we can assume that the growing personal preference to use Russian language coincided with the increase of new users. 
Some of these new users were likely to be people who joined the Facebook page after the end of the protest to follow the political agenda. It also could be that some of the new users were present on the page before, but they were silent. For some reason, they decided to start posting after the end of the Euromaidan. For example, it is possible that some users were quite busy following the events of the protest, and they did not have time or need to discuss political news. After the end of the protest, these people were no longer involved in the logistics, and they had more free time to debate political news. Finally, one could also assume that it was an influx of Russian bots or pro-Russian activists who entered this group to shape its content and nudge the discussion. We cannot quantify all these groups. Nevertheless, it is clear that the very composition of the group correlated with the personal usage of the Russian language. More than a half of \textit{active users} decided to use Russian more often after the end of the Euromaidan, precisely when the influx of new users happened. This finding indicates that online language behaviour is likely to be affected by the very structure of online communication (and not only personal preferences or national sentiments). 

\subsection{Local influence: If the language of posts shape the comment language preference of the active users?}

In our previous analysis, we investigated the possible influence of all users in the group and the influx of new users on the language behaviour of active users and referred to this as a \textbf{Global influence}. 
In contrast, in this part of the analysis, we move from the group level to the post level. More precisely, we analyze the role of the language of the post and the comments inside it and how they influence the choice of the comment language of other users, including \textit{active users}. We refer to it as \textbf{Local influence}.

\begin{figure}[t!]
\centering
\includegraphics[width=\columnwidth]{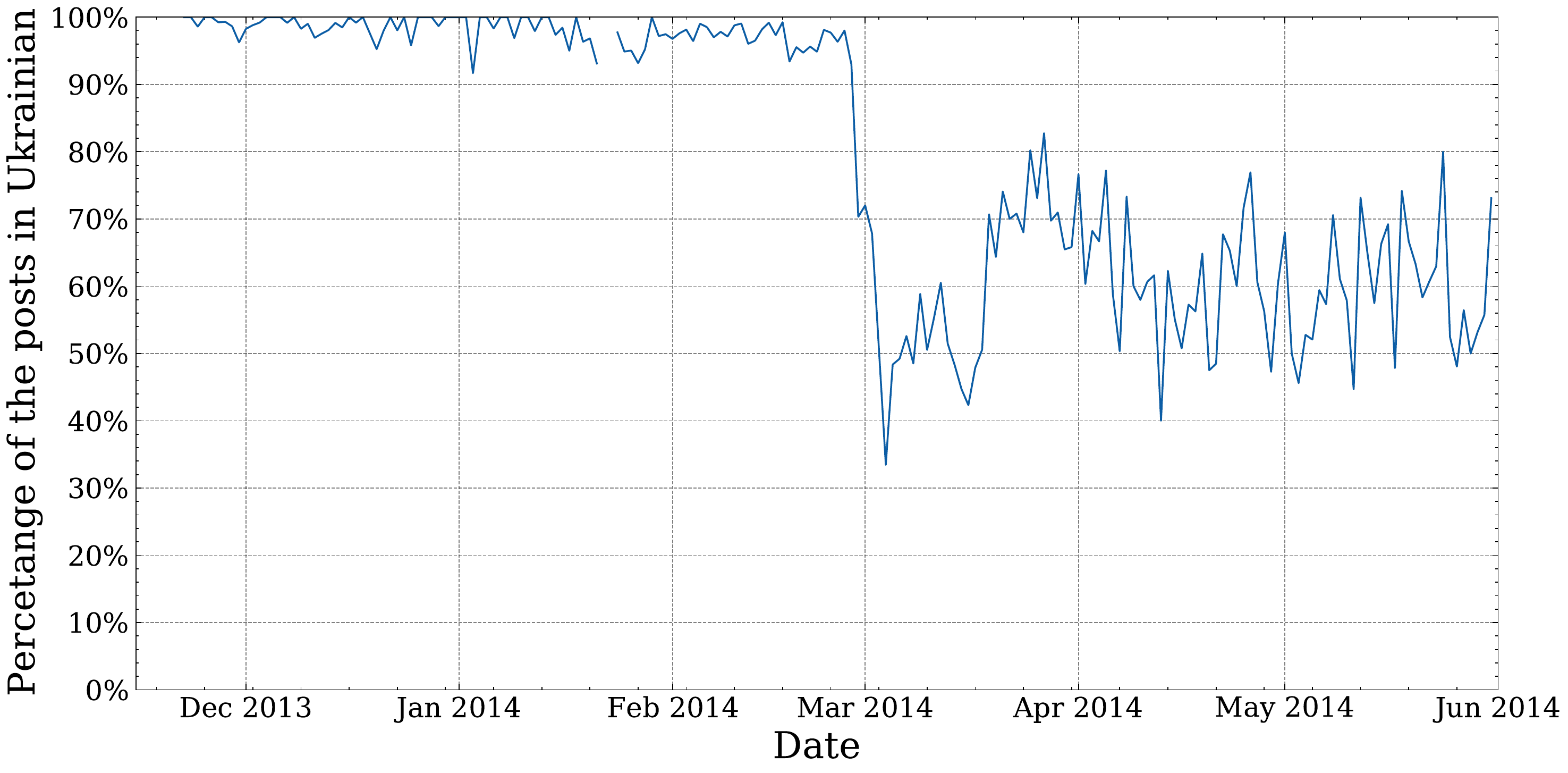}
\caption{Percentage of the posts in Ukrainian language}
\label{figure:percentage_in_ukrainian}
\end{figure}

\begin{figure}[t!]
\centering
\includegraphics[width=\columnwidth]{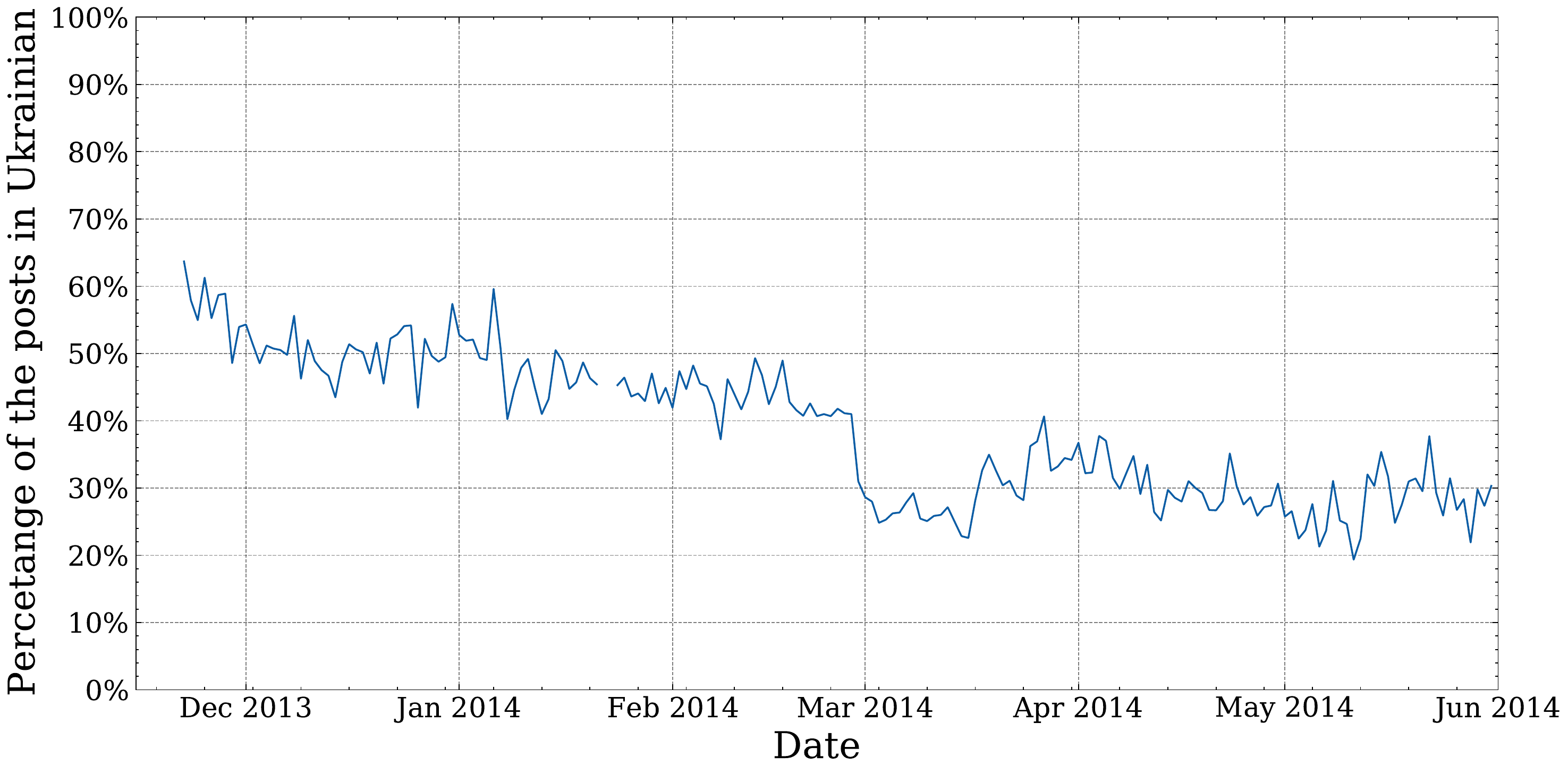}
\caption{Average of the fractions of the comments in the Ukrainian language per post}
\label{figure:mean_of_means_by_post}
\end{figure}

It could be that the administrators of the page increased the number of posts in Russian, thus provoking conversations in Russian. Figure \ref{figure:percentage_in_ukrainian} shows that, indeed, the percentage of Ukrainian posts dropped down after the end of the Euromaidan from almost 100\% to roughly 60 - 70 \%, meaning that for some reason administrators or moderators decided to post in the Russian language much more frequently. 
As well as the Figure \ref{figure:mean_of_means_by_post} that shows a declining trend of the fraction of comments in the Ukrainian language per post from roughly 50\% to 30\%. In other words, it shows the prevalence of the Russian language not only on a global level as was noticed before, but also all the posts were affected by that, so it had an effect on the local level. 

Nevertheless, we do not find evidence that the post language correlates with the language of the comment. Firstly, we check the Pearson's correlation coefficient for the general population of users between the language post and comment. The correlation coefficient is 0.15, which does not indicate a strong relationship, showing that users who leave comments do not mind the post language agenda. Then we check the same value but for the two separate samples: comments and posts before the end of Euromaidan and after. The coefficients are 0.05 and 0.12, respectively, which again do not show any strong evidence of the possible correlation. Finally, we test the same hypotheses on the sample of the \textit{active users} (with a threshold of 10 for both Russian and Ukrainian comments). It turns out that in this case, the Pearson correlation coefficient is 0.16 for the whole period. Before and after the split date, the coefficients are 0.06 and 0.15 respectively.

Therefore, we conclude that the language preferences of the \textit{active users} and the general population of users in our dataset cannot be easily nudged or affected by the post. This finding is in line with surveys that showed that Ukrainians were not likely to modify their prior preferences \cite{pop2018identity}.

\subsection{If the Russian language is more sticky than the Ukrainian language?}

Finally, in the last part of our analysis, we concentrate on the following question: why do we observe the significant evidence in the change of the usage of the Russian language but not for the Ukrainian language? To investigate this and provide a possible explanation, we look closer at the patterns of the language switches and compare how often the \textit{active users} (with threshold 10) switch from Russian to Ukrainian and vice versa. At the same time, we analyze users that use one language much more frequently than another, and we call them as \textit{loyal users} to their preferred language. Here, we define \textit{loyal users} as users who used a particular language in at least 80\% of all their comments. We also tried different thresholds (from 60\% to 90\% and received similar results, so we just report results only for the threshold equal to 80\%). This results in approximately 16\% the Ukrainian \textit{loyal users} and 10\% the Russian \textit{loyal users}, out of the total users, selected as \textit{active users}. 

Our first step is to reorganize the language vectors for each user in the following way. If the next language of the comment (ordered by time) is the same as the language of the current one, we put 0 (failure), otherwise 1 (success). After that, we analyze the normalized frequencies of such changes and the lengths of the longest sequences without switches.

\begin{figure*}[t]
	\subfloat[]{
		\includegraphics[width=0.45\textwidth]{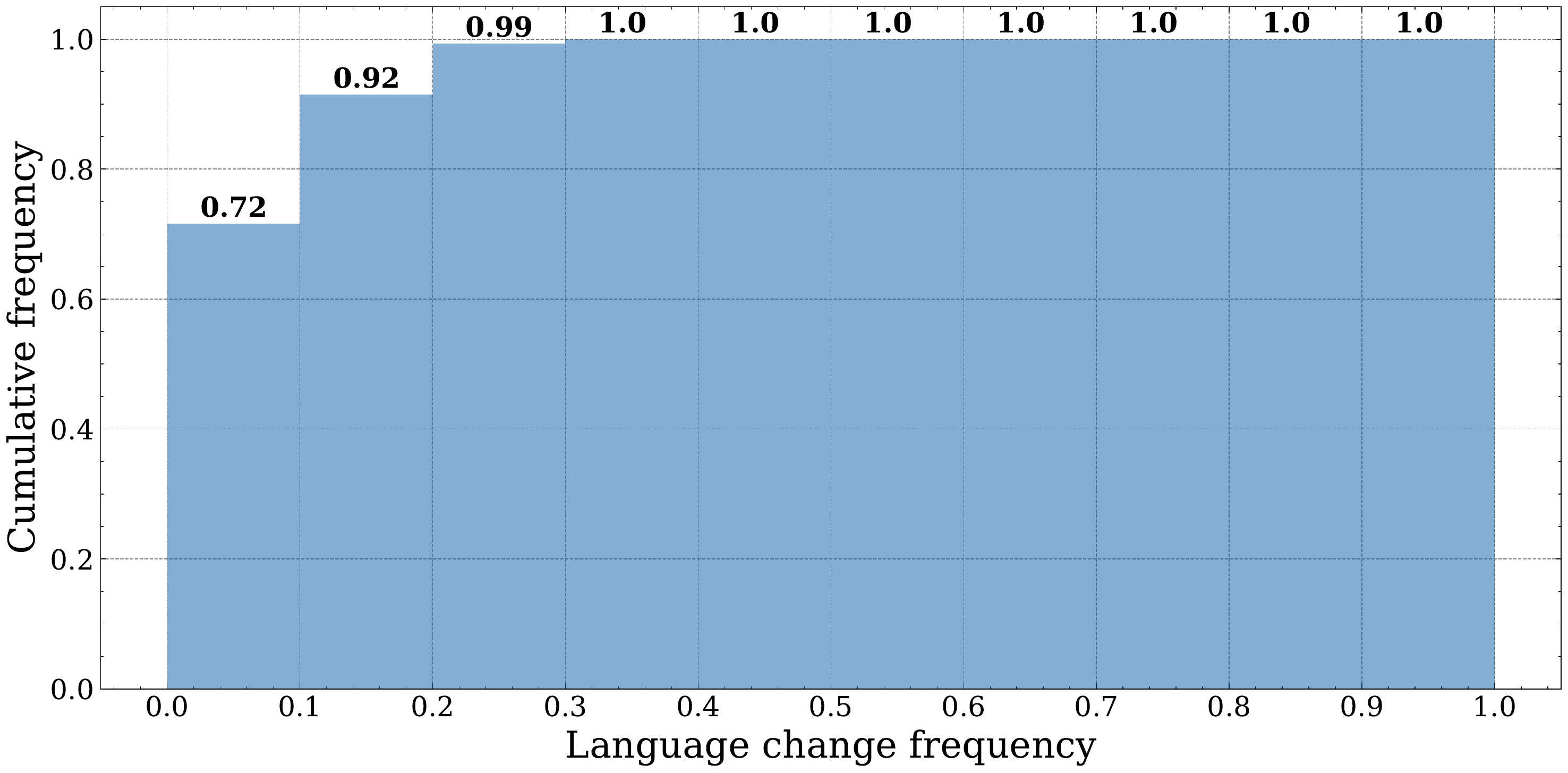}
		\label{plot:cumulative-russian}
	}~\hfill
	\subfloat[]{
		\includegraphics[width=0.45\textwidth]{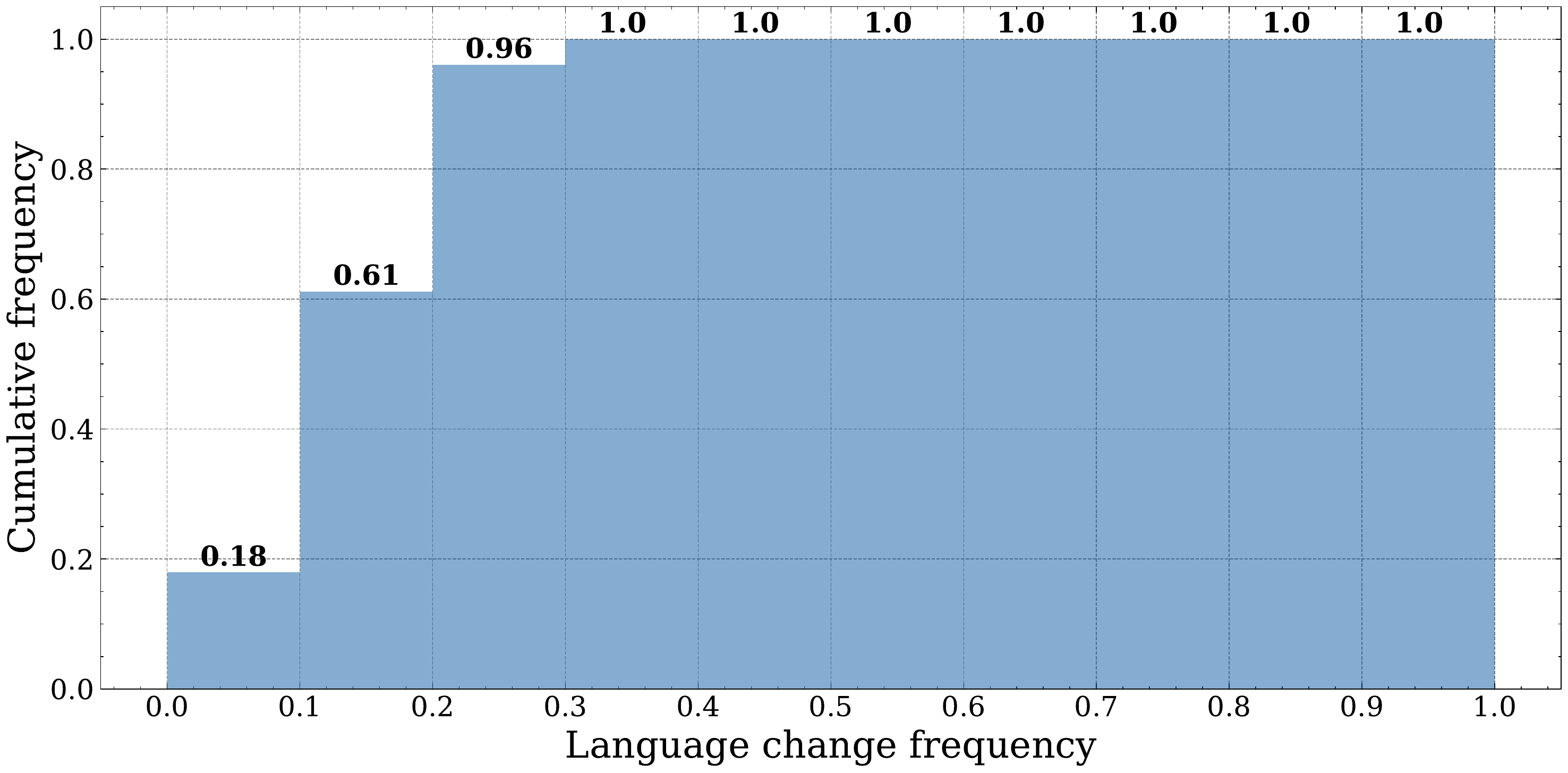}
		\label{plot:cumulative-ukrainian}
	}
\caption{Cumulative language change frequency for the Russian (a) and the Ukrainian (a) loyal users}
    \label{plot:cumulative-russian-ukrainian}
\end{figure*}

Then we analyze how frequently the users switch from one language into another. The comparison between the Russian and Ukrainian \textit{loyal users} is presented in Figure \ref{plot:cumulative-russian-ukrainian} (a) and (b), in turn. These figures show a remarkable difference between the Russian \textit{loyal users} and Ukrainian \textit{loyal users}. Russian \textit{loyal users} almost do not change their language behaviour in 72\% of comments, whereas Ukrainian \textit{loyal users} almost do not change their behaviour in only 18\% comments. To put it simply, the Ukrainian \textit{loyal users} are more flexible. They use the Ukrainian language once, and then they change it to Russian, and then they go back to Ukrainian, and so on.  

These findings again well matched with what we have observed in Sections 4.2 and 5.1. 
It seems that the influx of new users who commented in Russian incentivized those users who initially preferred Ukrainian to start using Russian on many occasions. Thus, the Ukrainian \textit{loyal users} had to use Russian more often after the end of the Euromaidan.

\subsubsection{Conclusion: }
In this section we investigate the possible explanations of why:
\begin{enumerate}
    \item Users' language behaviour changed (as they started to use the Russian language more than they used before) 
    \item  Why is the proportion of people who started to use the Russian language significantly different from those who started to use the Ukrainian language?
\end{enumerate}

We showed that the general population of users started to use the Russian language more frequently. In addition, right after the Euromaidan revolution, there was an influx of new users who brought the Russian language with them. Additionally, the moderators started to use the Russian language more often in their posts.

However, we didn't find a strong relationship between the language of the post and the comment for the users we analyzed. We found out that the users who mostly prefer the Ukrainian language switch between the Russian and the Ukrainian languages more often, while those who prefer Russian are more sticky. 
These findings are in line with the hypothesis from the literature that Ukrainian activists use language strategically (i.e., to respond to Russian comments) instead of being mobilized ethnically or nationally  \cite{metzger2016tweeting}.

\section{Limitations}
\label{chapter:limitations}

Although our paper is the first to present individual changes in language behaviour during the Euromaidan protest, it still has some limitations. First of all, we acknowledge that our data are from a single Facebook page. Other researchers investigated many Twitter users or regional platforms \cite{metzger2016tweeting, bond201261}. Nevertheless, we chose this group because it was the largest Facebook page which was the most popular among political activists. Therefore, all trends are salient and pronounced there. Not to mention that it provides a significant number of observations. Second, our dataset does not allow us to disentangle all nuanced social and psychological mechanisms that drive human behaviour. Nevertheless, our data analysis sheds new light on individual-level trajectories of online users.

\section{Conclusion}

To sum up, our work is the first paper to investigate individual changes in language behaviour (instead of aggregated data) during the Euromaidan protest. We show that active Ukrainian users changed their language preferences and used the Russian language more often after the end of Euromaidan. By showing this, we provide additional evidence that protesters use language rather strategically depending on the context instead of reacting to an ethnic or national mobilization. We also presented different ways to make sense of our findings. Although we could not provide conclusive modelling of individual behaviour due to the limitations of our data, we were able to propose some data-driven explanations. First, we can rule out the influence of the post language because there was no significant correlation between the languages of posts and comments. Even though the proportion of posts in Russian increased dramatically, active users did not react to it by switching to identical language in their comments. Therefore, we suggest that administrators were not able to nudge the language behaviour of active users.

On the other hand, we observe the appearance of many new users who speak Russian in the data. It seems that active users reacted to new users by talking with them in Russian. We also observe that the Ukrainian language is less sticky than the Russian language. In other words, even loyal Ukrainian speakers who were active before the end of the Euromaidan, are more likely to switch between languages after the end of the protest to react to new users. This finding is in line with the idea that Ukrainian activists use language strategically. They switch between languages depending on circumstances to facilitate communication instead of being nudged by national or ethnic mobilization. Although this explanation has been proposed in the literature \cite{metzger2016tweeting}, our paper is first to confirm this idea with the longitudinal data. Furthermore, our findings are in line with the results of representative surveys which show that Ukrainians are more likely to shift their short-term behavior and identities to reflect their identities rather than modify their identities \cite{pop2018identity,pop2021protest}.

\printbibliography

\end{document}